%% file: main.tex
\documentclass[sigconf,natbib=false]{acmart}
\AtBeginDocument{%
  }

\setcopyright{acmlicensed}
\copyrightyear{2026}
\acmYear{2026}
\acmConference[PELE]{Seventh
Annual Workshop on A/B Testing and Platform-Enabled Learning Engineering
(PELE), Learning@Scale 2026}{June 27-- July 03,
2026}{Seoul, Republic of Korea}
\acmISBN{978-1-4503-XXXX-X/2026/06}

\RequirePackage[
  datamodel=acmdatamodel,
  style=acmnumeric,
  ]{biblatex}

\begin{document}

\title{A Case Study Reexamining the Cold-Start Problem in Knowledge Tracing Models and Implications for SafeInsights, an Education Research Infrastructure}

\author{Jiayi Zhang}
\email{jzhang31@wpi.edu}
\orcid{0000-0002-7334-4256}
\affiliation{%
  \institution{Worcester Polytechnic Institute}
  \city{Worcester}
  \state{MA}
  \country{United States}
}

\author{Ryan S. Baker}
\email{ryanshaunbaker@gmail.com}
\orcid{0000-0002-3051-3232}
\affiliation{%
  \institution{Adelaide University}
  \city{Adelaide}
  \state{SA}
  \country{Australia}
}

\author{Debshila Basu Mallick}
\email{debshila@rice.edu}
\orcid{0000-0002-0597-3528}
\affiliation{%
  \institution{Rice University}
  \city{Houston}
  \state{TX}
  \country{United States}
}

\author{Cristina Heffernan}
\email{cristina.heffernan@assistments.org}
\orcid{0009-0009-4114-9892}
\affiliation{%
  \institution{ASSISTments}
  \city{Worcester}
  \state{MA}
  \country{United States}
}

\author{Neil Heffernan}
\email{nth@wpi.edu}
\orcid{000-0002-3280-288X}
\affiliation{
  \institution{Worcester Polytechnic Institute}
  \city{Worcester}
  \state{MA}
  \country{United States}
}

\renewcommand{\shortauthors}{Zhang et al.}
\renewcommand{\shorttitle}{Cold-Start Problem in KT Models and Implications for SafeInsights}


\begin{abstract}
Knowledge tracing (KT) models are widely used to predict students’ evolving knowledge states from their learning history. However, many KT models are evaluated using specific datasets, platforms, and learning contexts, raising questions about whether reported model performance replicates and generalizes across newer datasets that vary in context. This paper replicates and extends Zhang et al. (2021)\cite{zhang2021knowledge}, which examined the cold-start problem in KT models and found that deep-learning-based KT models performed better partly because of stronger predictions when students began practicing a skill. Using a more recent ASSISTments dataset, FoundationalASSIST, we replicate the previous analysis by evaluating model performance across opportunities to practice and extend the analysis by examining performance across problem types, including fill-in-the-blank, multiple-choice select-one, multiple-choice select-all, and order/sort problems. Results show that KT model performance varies across both student practice trajectories and problem types. Beyond the empirical replication, this study identifies practical challenges in reproducing educational data mining studies and serves as a proof of concept, showing how privacy-preserving research infrastructures such as SafeInsights can be leveraged to facilitate educational research and support replication analyses.
\end{abstract}

\ccsdesc[500]{Applied computing~Interactive learning environments}

\keywords{Knowledge Tracing, Data Enclave, SafeInsights, Open Datasets}

\maketitle

\input{Intro}

\input{Methods}

\input{Results}

\input{Discussion}

\begin{acks}
\begin{sloppypar}
We would like to thank Andrés Felipe Zambrano and Eamon Worden for their support of this study. 

This material is based upon work supported by the National Science Foundation under Award No. 2153481. Any opinions, findings, and conclusions or recommendations expressed in this material are those of the author(s) and do not necessarily reflect the views of the National Science Foundation.

Neil and Cristina Heffernan would also like to thank their many past and current funders including NSF (2118725, 2118904, 1950683, 1917808, 1931523, 1940236, 1917713, 1903304, 1822830, 1759229, 1724889, 1636782, 1535428, 2215842, 2225091, 2341948, \& 2153481); IES (R305N210049, R305D210031, R305A170137, R305A170243, R305A180401, R305A120125, R305R220012, R305T240029, \& R305J250058); DOE (U411B190024, \& S411A240012); GAANN (P200A120238, P200A180088), Schmidt Futures, BMGF, CZI, Arnold, Hewlett, the Jaffe Foundation, and anonymous donors. None of the opinions expressed here are those of the funders. 
\end{sloppypar}
\end{acks}

\printbibliography
\end{document}

%% file: Intro.tex
\section{Introduction}
\subsection{Open Datasets for Educational Data Mining}
In educational data mining (EDM), large-scale datasets collected from digital learning platforms (DLP) have played a central role in advancing the field. The most widely used datasets document students’ interactions with learning systems, including their responses to problems, timing, sequences of actions, and learning outcomes \cite{mihaescu2021review, svabensky2026open}. Over the past several decades, open-access datasets such as datasets from ASSISTments \cite{heffernan2014assistments,worden2026foundationalassist}, Datashop \cite{koedinger2010datashop}, Junyi Academy \cite{choi2020ednet}, and other large-scale student response datasets (e.g. \cite{wang2020diagnostic}) have enabled researchers to develop, compare, and improve computational models, measuring a range of learning behaviors and constructs, such as knowledge and mastery \cite{gervet2020deep,pelanek2017bayesian}, engagement and affect \cite{fancsali2014causal,walonoski2006detection}, and cognitive and metacognitive processes \cite{winne2000measuring}. As a result, they have been widely used to understand student learning processes, evaluate instructional supports \cite{feng2023implementing}, and train predictive models that estimate how and how well students learn under specific conditions, with some of the models implemented in the systems to inform adaptive supports (e.g., \cite{li2024flora}). 

\subsection{Challenge of Developing Generalizable Models}
Although these datasets have enabled substantial methodological progress, many models and empirical findings are developed or derived from one specific dataset, platform, student population, time period, or instructional context. This raises an important question for the field: how well do learner models trained in one context transfer and generalize to another? A model that performs well on the original dataset may not work equally well in a different platform, year, student population, or instructional context \cite{baker2019challenges}. This issue is especially important when the models are used to inform pedagogical changes or interventions, as they may lead to unintended effects if they have not been fully vetted in the new context, where key phenomena around engagement or self-regulated learning may manifest differently.

Given this concern, several studies have examined the transferability of models across platforms and populations. For example, \cite{paquette2015cross} analyzed the transferability of gaming the system models across platforms and modeling approaches. They found that both platform design and modeling approach influenced the level of transferability. Similarly, \cite{zhang2024using} and \cite{borchers2024using} analyzed the transferability of four self-regulated learning detectors across platform designs, subject domains, and languages. These LLM-based detectors, which were designed to identify SRL processes in think-aloud utterances, were found to transfer successfully across platform designs and languages, but less successfully across domains. Additionally, \cite{jensen2019generalizability} evaluated the generalizability of affect detectors across usage and demographic clusters, and \cite{lee2025conceptdrift} examined the generalizability of two knowledge tracing models over time, in which they concluded that a drastic change in the user base may influence model performance.

While these efforts show transferability is possible, these studies are standalone, individual cases that evaluate the generalizability of a specific model or a set of them across limited contexts. As learning platforms increasingly support data-intensive research, there is a growing need for \textbf{systematic} efforts to examine whether prior results and models replicate and under what conditions and contexts they generalize. In order to evaluate these models systematically, open datasets are necessary.

However, despite growing initiatives to release datasets, most digital learning platform datasets are still not commonly shared for public use \cite{svabensky2026open}. One major barrier is the privacy and safety concern associated with personally identifiable information (PII) encoded in student-level data. Preparing data for public release often requires substantial time and resources to de-identify records, remove sensitive information, and ensure that students, teachers, schools cannot be re-identified. 
As a result, dataset releases are often delayed (even when they do occur), meaning that researchers are typically analyzing data collected years earlier rather than data that reflect current platform use, instructional designs, or student populations. This delay limits the relevance of findings for current design needs.

In addition, depending on how data are processed before release, public datasets may be incomplete, disconnected from important contextual information, or formatted in ways that differ across time and studies. These issues make it difficult to replicate prior analyses, compare findings across datasets, and evaluate whether results remain applicable in new contexts. Therefore, the field needs research infrastructure that can support rigorous analysis of educational platform data while reducing the need to publicly release sensitive student-level datasets \cite{gardner2018morf,datatotrust2025codesigning, trustbydesign2025fivesafes}.

\subsection{SafeInsights for Privacy-Preserving Educational Research}
One emerging example is SafeInsights \cite{datatotrust2025codesigning}, a privacy-preserving education research hub designed to enable large-scale learning research without exposing individual student-level data to researchers or sharing student data out from the site of origin \cite{trustbydesign2025fivesafes,datatotrust2025codesigning}. Student privacy sits at the core of SafeInsights’ design and implementation, secured through complementary layers of technical safeguards and governance controls. The infrastructure equips partner data organizations, including DLPs (such as ASSISTments), with secure data enclaves paired with a compute-to-data model. Rather than sending data to researchers, researchers encode their research questions as analysis code and submit it to the partner DLP, which reviews the code and runs it against data held within the enclave. To support researchers with this workflow, they are provided with rich documentation, synthetic example datasets, and other resources for developing and refining analyses prior to submission. Finally, aggregated results are reviewed by the DLP before being returned to the researcher. SafeInsights builds on the approach introduced by the MOOC Replication Framework project \cite{gardner2018morf} and continues to incorporate lessons from it as the researcher and partner DLP experiences evolve \cite{baker2025morf}.

This infrastructure creates new possibilities for conducting replication and generalization studies across large-scale datasets while reducing privacy and data-sharing risks and minimizing processing times and effort from the data providers.

\section{The Current Study}
In this paper, we aim to demonstrate the potential of SafeInsights for educational research by presenting an example use case. Given that the SafeInsights infrastructure is currently under development, this paper serves as a proof of concept, showing how this infrastructure can be used to facilitate educational research and support replication analysis by evaluating what knowledge tracing models work the best in what contexts.

Specifically, in the current study, we replicate and extend an analysis conducted by Zhang et al. (2021) \cite{zhang2021knowledge}, which examined the performance of knowledge tracing (KT) models over attempts (instances) using the ASSISTments 2009 dataset \cite{heffernan2014assistments}. That study showed that the stronger aggregate performance of deep-learning-based KT models, compared with more traditional models such as Bayesian Knowledge Tracing (BKT; \cite{corbett1994knowledge}) and Performance Factors Analysis (PFA; \cite{pavlik2009performance}), may be partly explained by their stronger performance on initial predictions. This finding reflects a cold-start issue for traditional models, underscoring a broader issue in KT model evaluation: better overall performance metrics can obscure when and why a model performs better. It is important for developers selecting a model for implementation to know what contexts it works better in, and why. 

In this study, we replicate this prior analysis using FoundationalASSIST, a newer open dataset released by ASSISTments \cite{worden2026foundationalassist}. This dataset, even though publicly available, it serves as a test case for developing a standardized analytical workflow that could later be submitted to SafeInsights. 

By repeating the analytical workflow on a newer dataset, we examine whether the conclusion from Zhang et al. (2021) \cite{zhang2021knowledge} holds across datasets and contexts. In addition to replicating the prior analysis, we extend it by examining whether KT model performance varies by problem type, new contextual information included in the new dataset. Specifically, we compare model performance across multiple-choice select-one, multiple-choice select-all, fill-in-the-blank, and sort/order questions. This extension is motivated by the broader question of what model works best in what context. If KT models perform differently across problem types, then aggregated model comparisons may hide meaningful variation in when a model is most useful. Examining performance by problem type, therefore, provides a more contextualized understanding of KT model evaluation. Lastly, and broadly speaking, this study serves as an example for interacting with SafeInsights, demonstrating the process of developing a standardized analytical workflow that could potentially be submitted to a secure enclave to support future privacy-preserving replication studies.

As such, this paper aims to address the following research questions:

\begin{itemize}
    \item \textbf{RQ1:} Does the finding from Zhang et al. (2021) \cite{zhang2021knowledge} replicate? Specifically, do deep-learning-based knowledge tracing models perform better partly because of stronger performance on initial predictions?
    
    \item \textbf{RQ2:} Does the performance of knowledge tracing models vary by problem type?
    \end{itemize}
    
Through the process of conducting this replication and extension, what can we learn about practices that improve the feasibility of replicating EDM studies, especially in ways that support privacy-preserving infrastructures such as SafeInsights?

%% file: Methods.tex
\section{Methods}
\subsection{Learning Platform ad Datasets}
ASSISTments is a digital learning platform that supports K–12 students’ mathematics learning by allowing teachers to assign online math problems. The platform provides students with practice opportunities, immediate feedback, and on-demand support as they work through problems. ASSISTments has released several datasets that have been widely used in EDM and knowledge tracing research, including ASSISTments 2009 \cite{heffernan2014assistments}, ASSISTments 2013 \cite{san2013towards}, and FoundationalASSIST, the most recent release \cite{worden2026foundationalassist}.

These datasets include similar elements documenting student performance on math problems. Specifically, for each problem, the datasets report student ID, problem ID, associated skill, and whether the problem was answered correctly. To further facilitate KT research, FoundationalASSIST includes additional data elements, including problem bodies, problem type, and students’ actual responses \cite{worden2026foundationalassist}.

Even though the data elements are similar across ASSISTments 2009 and FoundationalASSIST, the design of problem sets in the ASSISTments platform has changed over time. The ASSISTments 2009 dataset was collected from earlier Skill Builder problem sets, in which each problem set typically included problems targeting the same skill. Students advanced to a different problem set, typically focused on a different skill, once they answered three questions correctly in a row. In contrast, FoundationalASSIST includes data from student use of ASSISTments between 2019 and 2024, including data collected beyond Skill Builder problem sets. In these additional types of problem sets, teachers can create problem sets and adjust advancement rules. As a result, a problem set may contain problems targeting different skills, and students may need to complete all problems within a problem set before advancing.

Despite the similarity of the data elements included in ASSISTments 2009 and FoundationalASSIST, differences in platform design may introduce nuances that influence the performance of KT models. For one, students in FoundationalASSIST tend to have longer trajectories, meaning they have more instances or attempts on a skill, given the absence of a rule that allows students to advance after three consecutive correct responses in certain problem sets. This change may, to some degree, mitigate the attrition issue reported in previous literature, such as the number of students being reduced to one-half or one-third over the first eight instances, but it may also introduce variation in KT model performance.

In this study, we collected students’ correctness on each problem, the associated skills, and problem type from the FoundationalASSIST dataset. Problem types included multiple-choice select-one, multiple-choice select-all, fill-in-the-blank, and order/sort.

Similar to Zhang et al. (2021) \cite{zhang2021knowledge}, we selected the four most practiced skills in the dataset to compare KT models’ performance over instances. The four selected skills were: 1) Solve Unit Rate Problems, 2) Use Proportional Relationships to Solve Ratio and Percent Problems, 3) Solve Scale Drawing Problems, and 4) Write and Evaluate Numerical Expressions with Exponents. We then further subset the data to include only students who had worked on a skill at least eight times. Due to the longer trajectories in FoundationalASSIST, this step did not significantly reduce the sample size from the total 5,000 students included in the original dataset. In total, 4,743 students worked on at least eight problems on these skills, resulting in 390,509 student-problem interactions. On average, each student completed 26 problems per skill, indicating relatively long skill-level trajectories in the current dataset.

Table \ref{tab:skill_descriptive} reports the number of unique problems, students, student-problem interactions, and the percentage correct for the four skills. The number of students is the same across the first eight instances for each skill because of the filtering criterion. Table \ref{tab:problem_type_descriptive} reports the same metrics by problem type.

\begin{table}[t]
\centering
\caption{Descriptive statistics by skill}
\label{tab:skill_descriptive}
\small
\begin{tabular}{p{0.40\columnwidth}rrrr}
\toprule
\textbf{Skill} & 
\textbf{N} & 
\textbf{Stud.} & 
\textbf{Prob.} & 
\textbf{Corr.} \\
\midrule
Solve Unit Rate Problems (Skill: 130) 
& 97,594 & 3,658 & 146 & 66\% \\

Write \& Evaluate Numerical Expressions with Exponents (Skill: 156) 
& 74,503 & 2,915 & 74 & 72\% \\

Use Proportional Relationships to Solve Ratio \& Percent Problems (Skill: 198) 
& 131,530 & 3,333 & 198 & 54\% \\

Solve Scale Drawing Problems (Skill: 219) 
& 96,253 & 3,189 & 113 & 51\% \\
\bottomrule
\end{tabular}
\end{table}

\begin{table}[t]
\centering
\caption{Descriptive statistics by problem type}
\label{tab:problem_type_descriptive}
\small
\begin{tabular}{p{0.38\columnwidth}rrrr}
\toprule
\textbf{Problem type} & 
\textbf{N} & 
\textbf{Stud.} & 
\textbf{Prob.} & 
\textbf{Corr.} \\
\midrule
Fill-in-the-blank(s) 
& 289,763 & 4,743 & 397 & 59\% \\

Multiple Choice (select 1) 
& 71,432 & 4,713 & 91 & 69\% \\

Multiple Choice (select all) 
& 36,790 & 4,632 & 38 & 47\% \\

Order / Sort 
& 1,895 & 1,571 & 4 & 34\% \\
\bottomrule
\end{tabular}
\end{table}

\subsection{Construct Knowledge Tracing Models}
As the goal of this study is to replicate the previous analysis, we followed the steps reported in Zhang et al. (2021)\cite{zhang2021knowledge} as closely as possible using the new dataset. Specifically, we split the dataset into five folds, nesting the data by student. This ensured that all data from a given student were contained within a single fold, preventing the same student’s data from appearing in both the training and testing sets and reducing the risk of data leakage and overfitting.

We trained and evaluated four KT models, three of which were used in \cite{zhang2021knowledge}: Bayesian Knowledge Tracing (BKT; \cite{corbett1994knowledge}), Performance Factors Analysis (PFA; \cite{pavlik2009performance}), and Dynamic Key-Value Memory Networks (DKVMN; \cite{zhang2017dynamic}). We additionally included Deep Knowledge Tracing (DKT; \cite{piech2015deep}) as an additional deep-learning-based model to extend the original comparison. The Python scripts for implementing the four models are available on GitHub.\footnote{\url{https://github.com/JZ2655/FoL_PELE2026_SafeInsights.git}}

\textbf{Bayesian Knowledge Tracing (BKT)} is a probabilistic model that estimates whether a student has mastered a skill based on their prior performance. For each skill, BKT assumes that students transition from an unmastered to a mastered state over time. The model updates the estimated probability of mastery after each student response while accounting for the possibility that students may answer correctly despite not knowing the skill (guessed), or answer incorrectly despite having mastered it (slipped). Because BKT models learning at the skill level using a small set of interpretable parameters, and is widely used in practice, it has been widely used as a traditional baseline in knowledge tracing research.

\textbf{Performance Factors Analysis (PFA)} is a logistic regression-based knowledge tracing model that predicts student correctness using students’ prior opportunities to practice a skill. Rather than explicitly modeling latent mastery states, PFA represents learning through accumulated prior successes and failures. Specifically, the model estimates the probability of a correct response based on the number of previous correct and incorrect attempts a student has made on the relevant skill. Compared with BKT, PFA provides a simpler and more directly observable representation of practice history, while still capturing how prior performance is associated with future correctness.

\textbf{Deep Knowledge Tracing (DKT)} is a deep-learning-based model that uses recurrent neural networks to model students’ sequences of responses. Unlike BKT and PFA, which rely on predefined skill-level assumptions, DKT learns sequential patterns from students’ interaction histories and uses these patterns to predict future correctness. This allows the model to capture more complex dependencies across skills and attempts. In the current study, DKT was included to extend the original comparison and examine whether another deep-learning-based KT model shows similar performance patterns across attempts and problem types.

\textbf{Dynamic Key-Value Memory Networks (DKVMN)} is another deep-learning-based KT model that uses a memory network architecture to represent student knowledge. In DKVMN, key memory captures latent concepts or skills, while value memory updates over time to represent a student’s evolving mastery of those concepts. This structure allows the model to track changes in student knowledge across sequences of interactions while preserving a more explicit representation of skill-related memory than standard recurrent models.

\subsection{Analysis}
As described above, and in alignment with Zhang et al. (2021) \cite{zhang2021knowledge}, we trained and evaluated the KT models using five-fold student-level cross-validation. For each test fold, we recorded each model’s predicted probability of correctness for each student-problem interaction.

Using these predicted probabilities, we computed Area Under the Receiver Operating Characteristic Curve (AUC) for each model, skill, and instance. An instance refers to the ordered practice opportunity for a student on a given skill. For example, the first time a student practiced a skill was labeled as the first instance, and subsequent opportunities to practice the same skill were labeled as later instances. This analysis addressed RQ1 by examining whether model performance varied as students gained more practice opportunities on a skill.

In addition, we computed AUC for each model by problem type, and where sample size permitted, by skill-by-problem-type combination, to address RQ2. This analysis allowed us to examine whether KT model performance varied across different problem formats.

\subsection{Replication-Oriented Workflow}
To support the broader goal of developing analyses that could be run in a privacy-preserving infrastructure (i.e., SafeInsights), the analytical process was implemented as a standardized workflow. The scripts specified the required input fields, pre-processing steps, student-level fold assignment, model training procedures, and evaluation outputs. The workflow was designed to return aggregate performance metrics rather than student-level records. This design reflects the type of analysis that could potentially be submitted to a secure enclave, where protected student-level data remain inaccessible to researchers while approved outputs are returned.

%% file: Results.tex
\section{Results}
\subsection{RQ1: Does the finding from Zhang et al. (2021) \cite{zhang2021knowledge} replicate?}
\begin{figure}[htp]
    \centering
    \includegraphics[width=\linewidth]{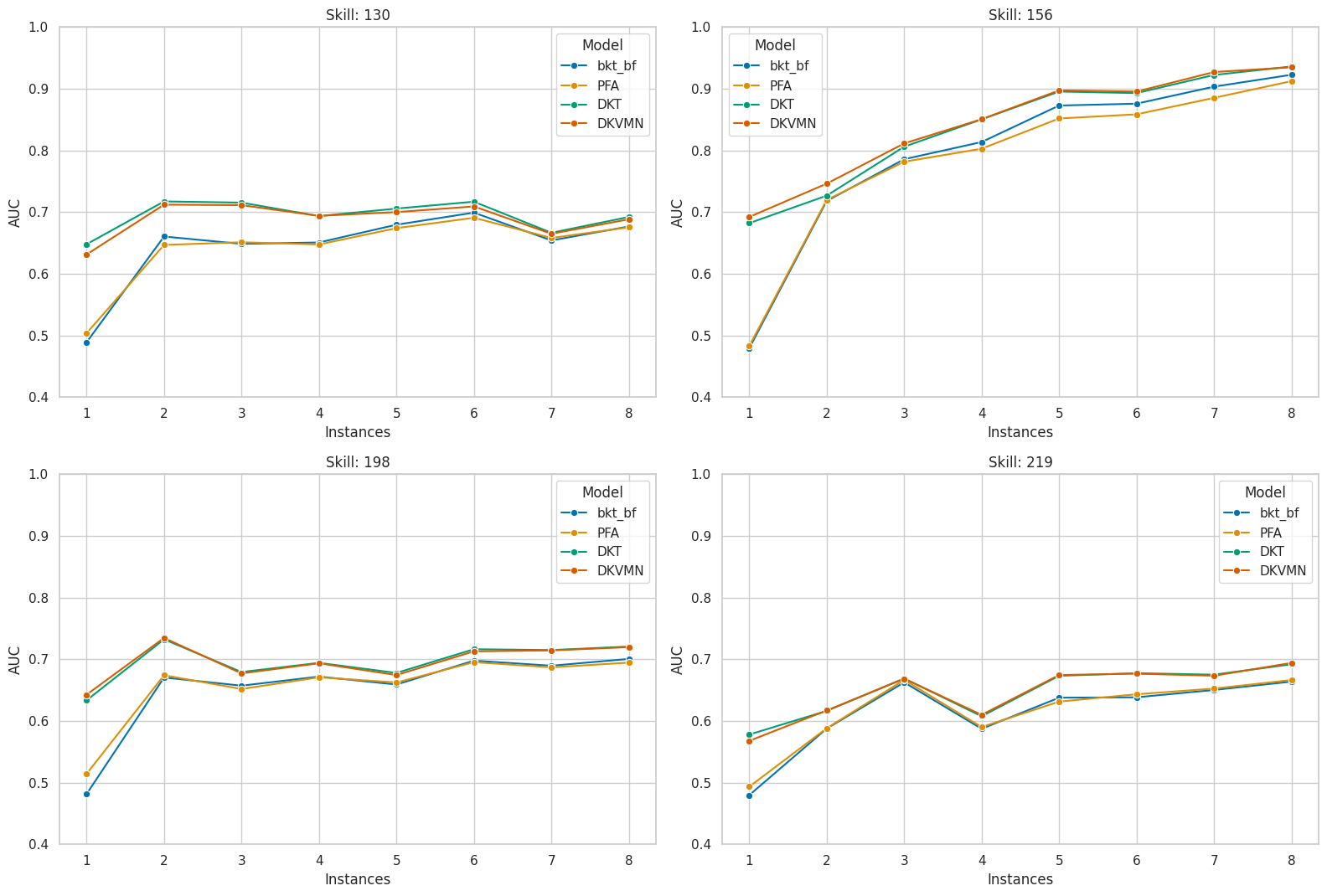}
    \caption{AUC by Skills, Models, and Instances}
    \label{fig:auc_by_instance}
\end{figure}

In Table \ref{tab:auc_models_instances}, we reported the average AUC for each of the first eight opportunities to practice (instances), for the four models across the four selected skills. Figure \ref{fig:auc_by_instance} shows the AUC across the first eight instances for each skill.

As shown in Table \ref{tab:auc_models_instances} and Figure \ref{fig:auc_by_instance}, all four models showed a general upward trend in AUC as students gained more opportunities to practice a skill. However, the difference between traditional and deep-learning-based models was mainly concentrated in the first two instances. At the first instance, DKT and DKVMN achieved AUC values of .635 and .633, respectively, whereas BKT and PFA achieved AUC values of .482 and .498. At the second instance, DKT and DKVMN continued to outperform BKT and PFA, but the gap was smaller.

After the first two instances, the performance of the four models became more similar. From Instances 3 to 8, all models followed similar trends, and the advantage of DKT and DKVMN over BKT and PFA was relatively small. By the eighth instance, BKT and PFA reached AUC values of .741 and .737, while DKT and DKVMN reached AUC values of .760 and .759. These results suggest that the stronger overall performance of deep-learning-based models is largely driven by their better performance during the initial attempts, when limited student history is available. 

Overall, this pattern is consistent with Zhang et al. (2021)\cite{zhang2021knowledge}; the findings replicate.

\begin{table}[t]
\centering
\caption{AUC by Models and Instances}
\label{tab:auc_models_instances}
\resizebox{\columnwidth}{!}{%
\begin{tabular}{lcccccccc}
\toprule
\textbf{Model} & \textbf{1} & \textbf{2} & \textbf{3} & \textbf{4} & 
\textbf{5} & \textbf{6} & \textbf{7} & \textbf{8} \\
\midrule
BKT   & .482 & .659 & .688 & .681 & .712 & .728 & .724 & .741 \\
PFA   & .498 & .657 & .688 & .678 & .705 & .722 & .721 & .737 \\
DKT   & .635 & .698 & .717 & .711 & .738 & .751 & .745 & .760 \\
DKVMN & .633 & .702 & .717 & .712 & .737 & .749 & .745 & .759 \\
\bottomrule
\end{tabular}%
}
\end{table}

\subsection{RQ2: Does the performance of knowledge tracing models vary by problem type?}
\begin{figure}[htp]
    \centering
    \includegraphics[width=\linewidth]{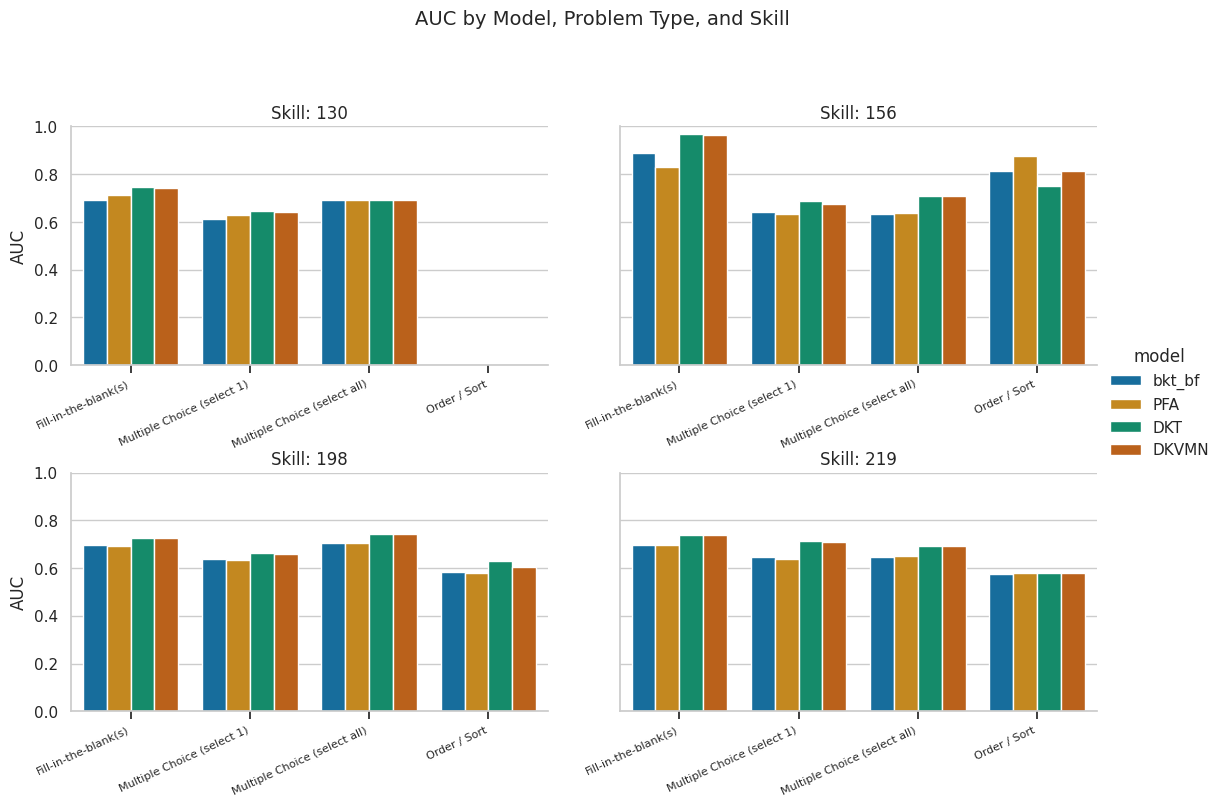}
    \caption{AUC by Skills, Models, and Problem Types}
    \label{fig:auc_by_problemType}
\end{figure}

As shown in Table \ref{tab:auc_models_problem_types} and Figure \ref{fig:auc_by_problemType}, model performance varied by problem type. Overall, AUC was consistently higher for fill-in-the-blank problems than for the other three problem types across the four KT models. 

Across problem types, DKT and DKVMN generally outperformed BKT and PFA. However, the size of this advantage differed by problem type. The advantage of deep-learning-based models was more noticeable for fill-in-the-blank and multiple choice select-all problems, where all four models performed better overall. This suggests that KT model performance is not only influenced by the number of prior attempts, as shown in RQ1, but also by the format of the problem being predicted.

\begin{table}[t]
\centering
\caption{AUC by Models and Problem Types}
\label{tab:auc_models_problem_types}
\resizebox{\columnwidth}{!}{%
\begin{tabular}{lcccc}
\toprule
\textbf{Model} & 
\textbf{Fill-in-blank} & 
\textbf{MC-1} & 
\textbf{MC-all} & 
\textbf{Order/Sort} \\
\midrule
BKT   & .768 & .654 & .656 & .568 \\
PFA   & .740 & .654 & .653 & .579 \\
DKT   & .808 & .695 & .709 & .593 \\
DKVMN & .807 & .692 & .709 & .591 \\
\bottomrule
\end{tabular}%
}
\end{table}

%% file: Discussion.tex
\section{Discussion and Conclusion}
\subsection{Contextualizing Knowledge Tracing Model Performance}

In this study, we show the cold-start pattern identified in \cite{zhang2021knowledge} was maintained in FoundationalASSIST, a dataset was collected in a different time period and includes additional way of how a problem sets can be created and used, resulting in different data characteristics. Specifically, the advantage of deep-learning-based KT models was mainly observed in the first two instances, when students had limited prior history on a skill. After students gained more practice opportunities, the difference between traditional and deep-learning-based models became smaller. As concluded in \cite{zhang2021knowledge}, this finding suggests the cold-start problem with traditional KT models and raise the issue underscoring the stronger aggregate performance of deep-learning-based models is partly driven by their stronger performance during initial predictions.

In addition, we found that KT model performance varied by problem type. Across the four models, AUC was highest for fill-in-the-blank problems. This suggests that model performance can also be influenced by the type of problem being predicted. One possible explanation is that students may interact with different problem types in different ways. For example, multiple-choice problems may allow more guessing or gaming behaviors than fill-in-the-blank problems. As a result, correctness may reflect factors beyond knowledge alone, even though BKT explicitly accounts for guessing and slipping. Recent work has shown KT models performance tend to decrease especially for random guess behaviors\cite{liu2026measuring}. Overall, these findings suggest that KT model evaluation should move beyond aggregate comparisons and examine model performance across meaningful contexts, including practice opportunities and problem types.

\subsection{SafeInsights Infrastructure to Support Educational Data Mining Research}
As demonstrated in this study, there is a need to replicate existing EDM research to better understand whether learner models generalize across contexts. If KT model performance varies across instances and problem types, then the same analysis should be repeated in additional contexts, such as different curricula, student populations, cohorts, platforms, and additional problem types (e.g. open-ended problems). This effort can be better supported by infrastructure such as SafeInsights, where researcher workflows are intended to be reproducible and can be tailored for different scenarios and populations. 

Through this case study, we identified several practices that can make replication more feasible, offering points for data partners and future researchers who are interested in using SafeInsights to conduct studies. For data providers and partners, including datasets that are diverse in context can better support investigations of generalizability. Second, documentation is needed to clarify how variables are defined and the context in which the data were collected. For researchers interested in using SafeInsights to conduct replication analyses, standardized analytical scripts are needed to make pre-processing, model training, and evaluation decisions explicit. Through this effort, SafeInsights can support EDM research not only by expanding responsible access to protected data, but also by encouraging more reusable, transparent, and cumulative analyses across educational contexts.

\subsection{Limitations}
We acknowledge several limitations of the current study. First, the current study focused on a limited set of KT models and skills. Although the selected models represent both traditional and deep-learning-based KT approaches, other KT models may show different performance patterns across instances and problem types. Similarly, the analysis focused on the four most practiced skills to ensure sufficient data across attempts. As a result, the findings may not represent model performance on less frequently practiced skills or skills with shorter student trajectories. However, this limitations can be potentially explored in future studies conducted with SafeInsights.

Second, although the workflow was designed with SafeInsights-like infrastructure in mind, the current analysis was conducted on an open dataset and was not executed inside a secure enclave. Therefore, the study focuses on demonstrating the feasibility of developing a standardized replication workflow rather than evaluating the full SafeInsights infrastructure. Because the infrastructure is expected to be ready for deployment after the dissemination of this paper, we anticipate testing similar workflows directly within privacy-preserving infrastructures in the near future to examine how well they support script submission, data access control, result review, and reproducible analysis.

\subsection{Conclusion}
This study replicated and extended Zhang et al. (2021) \cite{zhang2021knowledge} using FoundationalASSIST to examine whether KT model performance varies across instances and problem types. The results suggest that the cold-start pattern identified in prior work was maintained in the newer dataset, with deep-learning-based KT models showing better performance in the initial instances. The findings also show that KT model performance varies by problem type, highlighting the need to evaluate learner models across meaningful contexts rather than relying only on aggregate metrics. More broadly, this study demonstrates how standardized analytical workflows, supported by infrastructures such as SafeInsights, can help advance more replicable, transparent, and privacy-preserving EDM research.